\documentclass[useAMS,usenatbib]{mn2e}
\usepackage{graphicx}
\usepackage{amssymb}
\usepackage{amsmath}
\usepackage{times}

\newcommand{\Ha}{H$\alpha$}
\newcommand{\Hb}{H$\beta$}

\newcommand{\Oiii}{[{\sc O$\,$iii}]}
\newcommand{\Nii}{[{\sc N$\,$ii}]}

\newcommand{\placefigone}{
\begin{figure*}
\begin{center}
\includegraphics[width=\textwidth]{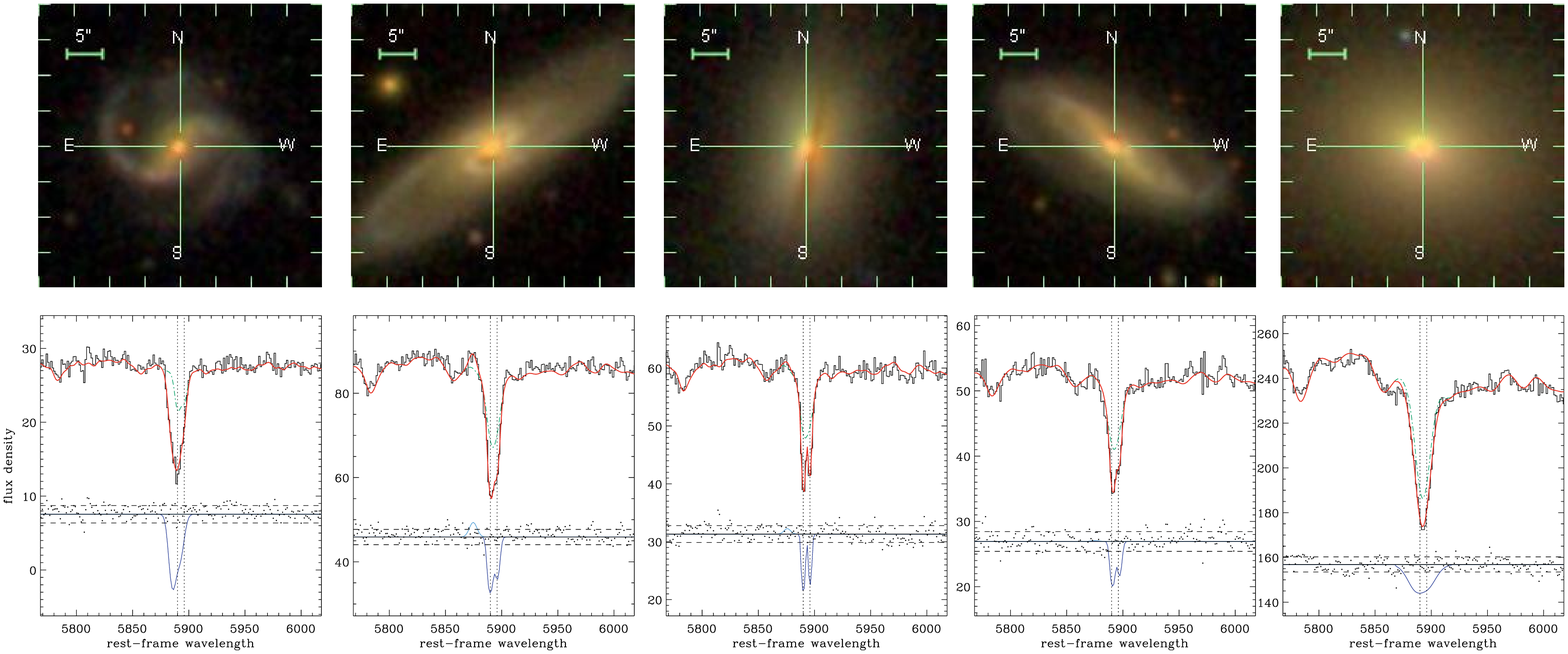}
\end{center}
\caption{SDSS colour images (top row) and spectral fits in the NaD
  region (bottom row) for five of our sample galaxies. In the lower
  panels the red line shows our best-fitting model, which is a
  superposition of our best {\tt GandALF\/} fit to the stellar
  continuum and nebular emission across the entire spectrum (green
  dashed and cyan lines, respectively) and of our fit to the NaD
  absorption excess using the formalism of \citet[][blue line, see
    \S~\ref{sec:analysis}]{Rup05}. The residuals of our fit (black
  dots) have been rescaled by adding a constant (marked by the
  horizontal lines), and their standard deviation (horizontal dashed
  lines) is used to assess the level of detection of the NaD excess
  features. The vertical dotted lines mark the rest-frame position of
  the NaD lines.
  }
\label{fig:one}
\end{figure*}
}

\newcommand{\placetabone}{
  \begin{table}
    \centering
    \renewcommand{\footnoterule}{}  
    \begin{tabular}{lcccc}
      \hline \hline
      \textbf{Morph. / em.-line class} & \multicolumn{2}{c}{\textbf{all objects}} & \multicolumn{2}{c}{\textbf{mJIVE-20 det.}}\\
      \hline
      & all & with NaD & all & with NaD \\
      \hline
      \hline
      ETG / SF &       19 &       6 &       1 &       0 \\ 
      ETG / TO &       15 &       6 &       4 &       1 \\
      ETG / Sy &       19 &       4 &       3 &       0 \\
      ETG / LI &        9 &       4 &       5 &       2 \\
      ETG / NE &      204 &      42 &      80 &      15 \\
      \hline
      LTG / SF &       69 &      29 &       0 &       0 \\ 
      LTG / TO &       46 &      26 &       2 &       0 \\
      LTG / Sy &       26 &       9 &       1 &       0 \\
      LTG / LI &        9 &       3 &       3 &       1 \\
      LTG / NE &       27 &      16 &       0 &       0 \\
      \hline
      Merger / SF &        4 &       1 &       0 &       0 \\ 
      Merger / TO &        0 &       0 &       0 &       0 \\
      Merger / Sy &        3 &       0 &       2 &       0 \\
      Merger / LI &        2 &       0 &       1 &       0 \\
      Merger / NE &        4 &       2 &       1 &       1 \\
      \hline
    \end{tabular}
    \caption[]{Breakdown of our sample according to galaxy morphology
      and nebular emission classification (where NE stands for objects
      with little or no emission), for all our objects and those with
      a VLBI radio core, as well as for those with detected NaD
      excess.}
    \label{tab:one}
  \end{table}
}

\newcommand{\placefigtwo}{
\begin{figure*}
\begin{center}
\includegraphics[width=\textwidth]{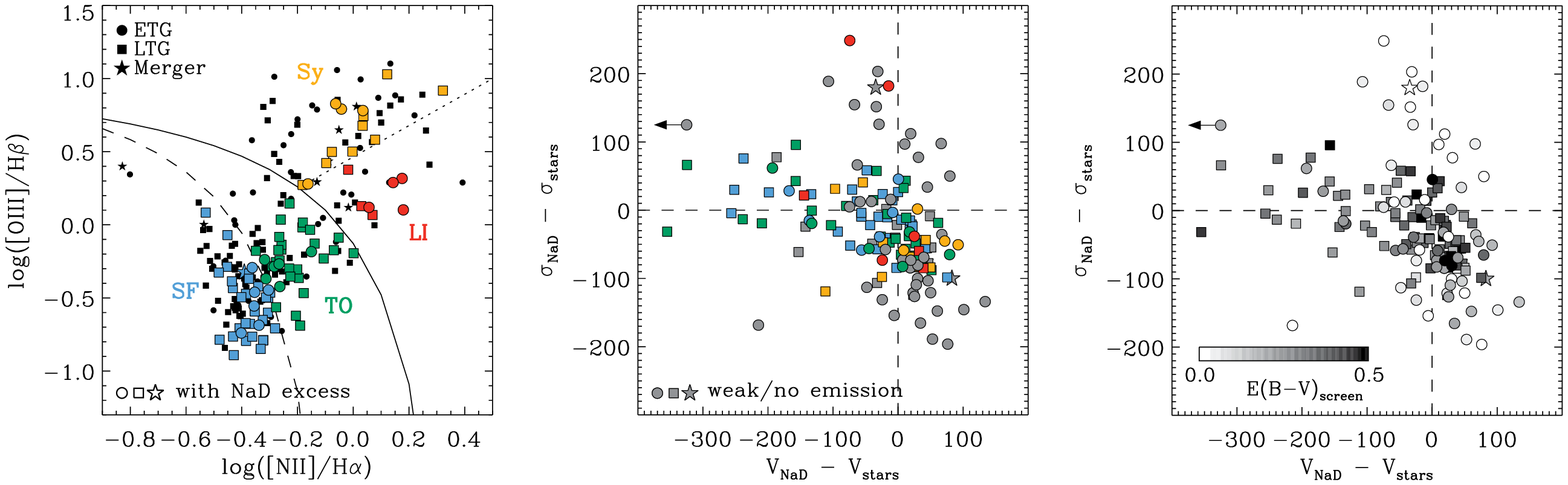}
\end{center}
\caption{Emission-line classification and NaD kinematic properties of
  our sample galaxies. Left: \Nii/\Ha\ vs. \Oiii/\Hb\ BPT diagram for
  mJIVE-20 targets with noticeable nebular emission (i.e., with $\rm
  A/N>3$ in all four lines). The shape of the symbols shows the galaxy
  morphology, whereas coloured symbols highlight objects with
  significant NaD absorption excess (i.e., with A/N for NaD $<-3$) and
  their emission-line classification. The dashed, solid and dotted
  lines showing the boundary between objects with star-forming,
  composite, Seyfert and LINER emission are from \citet{Kau03},
  \citet{Kew01} and \citet{Sch07}, respectively.
  Middle and right: $V_{\rm NaD}-V_{\rm stars}$ vs. $\sigma_{\rm
    NaD}-\sigma_{\rm stars}$ velocity and velocity dispersion offset
  diagrams, showing the kinematic properties of the NaD lines for
  objects that show a significant NaD absorption excess. In the middle
  panel the symbols are colour-coded according to the emission-line
  classification or are shown in grey if nebular emission was weak or
  absent. In the right panel darker shades of grey for the symbols
  correspond to increasing amounts of reddening by dust (affecting the
  entire SDSS spectra).
  Objects with significant amount of dust extinction display an NaD
  kinematics consistent with outflows (for negative $V_{\rm
    NaD}-V_{\rm stars}$ values), settled dusty disks (with narrower
  interstellar NaD line profiles w.r.t stellar absorption features),
  and possibly also mild inflows. On the other hand, in galaxies
  without much reddening any NaD excess is more likely associated to
  template-mismatch effects, and these objects tend to show rather
  large $\sigma_{\rm NaD}-\sigma_{\rm stars}$ values.}
\label{fig:two}
\end{figure*}
}

\newcommand{\placefigthree}{
\begin{figure*}
\begin{center}
\includegraphics[width=\textwidth]{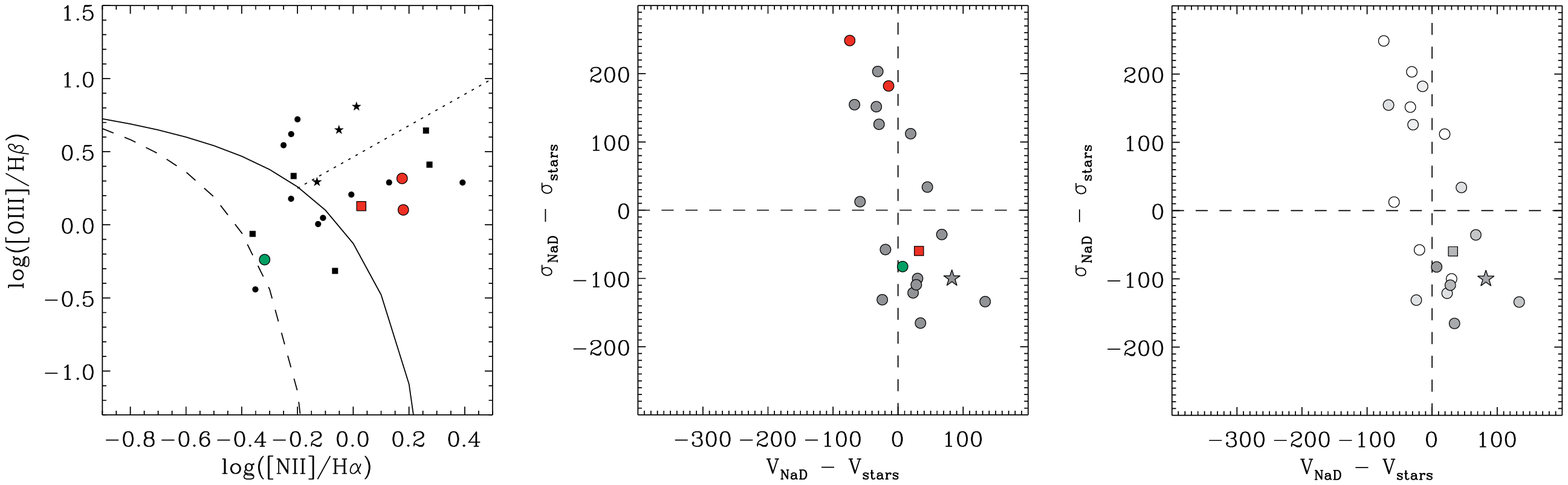}
\end{center}
\caption{Same as Fig.~\ref{fig:two} but now only for objects with
  detected radio cores in the mJIVE-20 survey. These are predominantly
  quiescent or weakly emitting early-type galaxies with NaD excess due
  either to the presence of settled disks and dust lanes, or which is
  induced by template-mismatch.}
\label{fig:three}
\end{figure*}
}

\title[NaD outflows in the mJIVE-20 survey]{Cold-gas outflows in
  typical low-redshift galaxies are driven by star formation, not AGN}

\author[M. Sarzi et al.]{\parbox{\textwidth}{Marc
    Sarzi$^{1}$\thanks{E-mail: m.sarzi@herts.ac.uk}, Sugata
    Kaviraj$^{1}$, Borislav Nedelchev$^{1}$, Joshua Tiffany$^{2}$,
    Stanislav S. Shabala$^{3}$, Adam T. Deller$^{4}$ and Enno
    Middelberg$^{5}$}\vspace{0.4cm}\\
\parbox{\textwidth}{
$^{1}$Centre for Astrophysics Research, University of Hertfordshire,
  College Lane, Hatfield AL10 9AB, United Kingdom\\
$^{2}$Astronomy Unit, School of Physics and Astronomy, Queen Mary
  University of London, London E1 4NS, UK \\
$^{3}$School of Mathematics and Physics, University of Tasmania,
  Private Bag 37, Hobart, TAS 7001, Australia\\
$^{4}$The Netherlands Institute for Radio Astronomy (ASTRON),
Dwingeloo, The Netherlands \\
$^{5}$Astronomisches Institut der Ruhr-Universitat Bochum,
Universitatsstrae 150, D-44801 Bochum, Germany} }

\begin{document}


\pagerange{\pageref{firstpage}--\pageref{lastpage}} \pubyear{2013}

\maketitle

\label{firstpage}

\begin{abstract}
Energetic feedback from active galactic nuclei (AGN) is an important
ingredient for regulating the star-formation history of galaxies in
models of galaxy formation, which makes it important to study how AGN
feedback actually occurs in practice.
In order to catch AGNs in the act of quenching star formation we have
used the interstellar Na{\sc$\,$i} $\lambda\lambda5890,5895$ (NaD)
absorption lines to look for cold-gas outflows in a sample of 456
nearby galaxies for which we could unambigously ascertain the presence
of radio AGN activity, thanks to radio imaging at milli-arcsecond
scales.
While compact radio emission indicating a radio AGN was found in 103
galaxies (23\% of the sample), and 23 objects (5\%) exhibited NaD
absorption-line kinematics suggestive of cold-gas outflows, not one
object showed evidence of a radio AGN and of a cold-gas outflow
simultaneously. Radio AGN activity was found predominantly in
early-type galaxies, while cold-gas outflows were mainly seen in
spiral galaxies with central star-formation or composite
star-formation/AGN activity. Optical AGNs also do not seem capable of
driving galactic winds in our sample.
Our work adds to a picture of the low-redshift Universe where cold-gas
outflows in massive galaxies are generally driven by star formation
and where radio-AGN activity occurs most often in systems in which the
gas reservoir has already been significantly depleted.
\end{abstract}


\section[]{Introduction}
\label{sec:intro}

Understanding the processes that quench star formation activity is
key to following how galaxies evolve over cosmic time.
In massive galaxies, which have deep potential wells, the current view
is that stellar winds and supernovae explosions are not powerful
enough to eject gas and regulate star formation
\citep[e.g.,][]{Sil98}.
Instead, this quenching is typically attributed to a central active
galactic nucleus \citep[AGN, e.g.,][]{Cro06,Sha09}, because
the energy released by the growth of the black hole can be orders of
magnitude larger than the binding energy of the gas, even in the most
massive systems \citep[e.g.,][]{Fab12}.
Observational evidence for this process remains mixed, however. While
some examples of AGN-driven outflows do exist both at high and low
redshifts \citep[e.g.,][respectively]{Nev08,Nyl13}, it is unclear
whether gas outflows are routinely driven by AGNs. In particular,
recent work at low redshift indicates that the onset of the AGN
appears to lag behind the peak of the starburst by several dynamical
timescales \citep{Kav09,Wil10,Sha12,Kav15b}, which implies that the
gas reservoir has been significantly depleted before the AGN had a
chance to couple to it.
To probe this issue in more detail, it is desirable to investigate
whether signatures of cold-gas outflows exist in systems that host
optical or radio AGNs and to see how these outflows compare to those
in star-bursting systems.

An effective way of carrying out this exercise is to explore the
presence of interstellar Na{\sc$\,$i} $\lambda\lambda5890,5895$ (NaD)
absorption in nearby galaxies.
When outflows occur in galaxies, neutral material entrained in the
flow can be observed against the stellar background via characteristic
absorption lines \citep[e.g.,][]{Rup05,Sat09,Che10}, particularly in
systems with low inclination. At optical wavelengths, the strongest of
such absorption lines are produced by neutral Sodium and observed via
the NaD doublet.
Thus, if AGN feedback routinely shuts down star formation by driving
cold-gas outflows, one expects to find an excess of NaD absorption
over what is expected from stellar photospheres alone, which is
blueshifted compared to the position of the photospheric NaD lines.

In this Letter, we use optical spectra from the Sloan Digital Sky
Survey \citep[SDSS, ][]{Aba09} to investigate the NaD properties of a
sample of galaxies that have been observed by the mJy Imaging VLBA
Exploration at 20\,cm survey \citep[][hereater mJIVE-20]{Del14}. While
SDSS spectra enable us to identify optical AGNs, the sub-arcsecond
radio imaging obtained with the Very Long Baseline Array (VLBA) during
the mJIVE-20 survey allows us to unequivocally identify the presence
of radio AGN activity. Taken together, these data enable a detailed
exploration of whether AGN are an important driver of outflows in the
massive galaxy population.

This paper is organised as follows. In \S~\ref{sec:data} we describe
our sample together with the mJIVE-20 survey and SDSS data that
underpin this study, whereas \S~\ref{sec:analysis} describes our
analysis of the NaD absorption lines. In \S~\ref{sec:results} we then
scrutinise the kinematics of the NaD lines to robustly identify
objects with interstellar NaD absorption and look for evidence for
starburst- or AGN-driven cold-gas outflows. Finally, we draw our
conclusions in \S~\ref{sec:conclusions}.

\section{Sample and Data}
\label{sec:data}

Our sample comprises of galaxies observed during the mJIVE-20 survey
for which images and spectra are available from the SDSS. 
mJIVE-20 is a survey that systematically observed objects detected by
the Faint Images of the Radio Sky at Twenty cm \citep[FIRST,
][]{Bec95}, with a median detection threshold of 1.2 mJy/beam and a
typical beam size of $\sim0\farcs01$ that effectively allows to zoom
in on the FIRST targets to further study the nature of the radio
emission and ascertain in particular the presence of a radio core
associated with an AGN.
The mJIVE-20 survey has targeted $\sim$25,000 FIRST sources, with
$\sim$ 5000 very-long baseline interferometry (VLBI) detections.

\placetabone
We cross-matched the mJIVE-20 targets with the DR7 value-added SDSS
catalogue of \citet[][hereafter OSSY]{Oh11}, since this contains
spectral measurements for the stellar kinematics and the nebular
emission within the 3\arcsec\ SDSS fiber. Based on {\tt pPXF\/} and
{\tt GandALF\/} fits \citep{Cap04,Sar06}, the OSSY catalogue also
provides us with a measurement for the level of detection of the
emission lines through their A/N ratio, between the peak amplitude of
the lines and noise level in the {\tt GandALF\/} fit residuals, and a
gauge for the amount of dust extinction that affects the entire SDSS
spectrum. Such a diffuse E(B-V) measurement is particularly relevant
to this work, since it corresponds well to the presence of
interstellar NaD absorption \citep{Jeo13}.
The OSSY database also readily provides us with the best-fitting model
for the stellar continuum, which is central to correctly characterise
the extent and kinematics of interstellar NaD absorption.
Finally, the OSSY catalogue is restricted to targets at $z<0.2$ so
that a robust visual classification of the galaxy morphology is
possible.

Overall, our sample includes 456 galaxies with a median stellar mass
of $10^{11}M_{\odot}$. Of these, 58\% are early-type galaxies (ETGs),
39\% are spirals (LTGs) and 3\% are on-going mergers.  Approximately
23\% of our sample (103 galaxies) have VLBI detections, 90\% of which
occur in ETGs.
Tab.~\ref{tab:one} provides a breakdown of our sample according to
galaxy morphology and to whether they have a VLBI radio core.

\section{Spectral Analysis}
\label{sec:analysis}

In order to assess the presence of interstellar NaD absorption and
measure the kinematics of such absorbing clouds, we follow the
approach of \citet{Rup05} after normalising the SDSS spectra for the
best-fitting stellar continuum provided by the OSSY catalogue.
We assume a single Maxwellian velocity distribution for the absorbers
along the line of sight and that the covering fraction $C_f$ of such
clouds is itself independent of velocity. Under these assumptions the
absorption profile of the interstellar NaD lines can be written as
\begin{multline}
I(\lambda) = 1 - C_f
\Big\{ 1 - \exp \Big[
-2\tau_0 e^{-(\lambda -\lambda_{\rm blue})^2
/(\lambda_{\rm blue} b / c)^2} \\
- \tau_0 e^{-(\lambda - \lambda_{\rm red})^2
/(\lambda_{\rm red} b / c)^2}
\Big] \Big\}
\end{multline}
where $\lambda_{\rm blue}$ and $\lambda_{\rm red}$ are the redshifted
central wavelength of the Na{\sc$\,$i} doublet, $\tau_0$ is the
optical depth at the centre of the red line, and $b=\sqrt{2}\sigma$ is
the Doppler parameter that measures the width of the lines. This
parametrisation thus yield a measurement for the velocity $V_{\rm
  NaD}$ and width $\sigma_{\rm NaD}$ of the NaD lines, as well as the
optical depth $\tau_0$ and covering factor $C_f$ of the absorbing
clouds.

After normalising the SDSS spectra of our sample galaxies for the OSSY
continuum model we proceed to fit with the \citeauthor{Rup05} method
any possible excess of NaD absorption in the data compared to what is
expected from the photospheric absorption provided by the OSSY best
fitting stellar-population model.
Our approach implicitly assumes that any NaD absorption excess is
due to interstellar absorption but in fact this is not always the
case.
Indeed, ETGs showing little or no dust absorption in their images
or nebular emission in their spectra often exhibit an NaD excess that
is most likely due to an enhanced [Na/Fe] abundance in their stellar
populations that cannot be accounted for by most stellar population models
\citep{Jeo13}, in particular given that in such smooth and ordinary
ETGs the NaD excess shows little or no kinematic offset compared to
the systemic velocity \citep{Par15}.
However, in what follows, we will show that objects where the NaD
excess stems from template mismatch can be clearly isolated in terms
of both the position and width of their best-fitted NaD profile in
addition to their reddening and nebular properties. 

\placefigone

Fig.~\ref{fig:one} shows SDSS colour images and our spectral fit in
the NaD region for five galaxies in our sample, all of which display a
significant excess of NaD absorption compared to the photospheric NaD
absorption of the best-fitting stellar-population model (i.e., the A/N
ratio for the fitted NaD profile is $<\!-3$).
Fig.~\ref{fig:one} also indicates the different nature of the
NaD absorption excess between spiral or dusty ETGs and quiescent
objects. Indeed, whereas in spiral galaxies and dust-lane ellipticals
our fit returns narrow NaD profiles that can be either blue- or even
slightly redshifted (first four panels from the left), the NaD excess
of quiescent ETGs (rightmost panel) is generally characterised by
rather broad NaD model profiles. Tab.~\ref{tab:one} also specifies
where NaD excess occurs in our sample.

\section{Results}
\label{sec:results}

\subsection{Entire Sample}
\label{sec:results_whole}

\placefigtwo
\placefigthree

We begin by discussing the emission-line classification and, where
present, the properties of the NaD excess for our entire sample.
Fig.~\ref{fig:two} (left panel) shows the standard
\citet[][BPT]{Bal81} diagnostic diagram that juxtaposes the
\Oiii/\Hb\ and \Nii/\Ha\ line ratios \citep{Vei87} for objects with
A/N$>3$ for all four lines, from which we deduce their emission-line
class. Of the 221 objects that are in this BPT diagram, which are
nearly 48\% of the sample (see also Tab.~\ref{tab:one}), most show
emission either dominated by star-forming regions (hereafter SF, 42\%
of active objects) or by the likely superposition of star-bursting and
AGN activity (TO for ``transition objects'', 28\%), with the remaining
objects exhibiting either Seyfert activity (Sy, 22\%) or central LINER
emission (LI, 9\%).
The remaining 235 objects in our sample (nearly 52\%) display only
weak (e.g., with only \Ha\ and \Nii\ detected emission) or no nebular
emission, which is puzzling considering that only 1/3 of these objects
show a compact radio core at VLBI resolution so that radio AGN
activity could account for their FIRST radio flux.
On the other hand, since 87\% of such quiescent or weakly emitting
objects are ETGs, it is likely that for many of those without a VLBI
detection the FIRST radio flux is due to circumnuclear or rather
extended star-formation activity, which would add to existing evidence
of recent star-formation in many ETGs based on UV imaging
\citep[e.g.,][]{Yi05,Jeo07}.

Moving on with the properties of the NaD absorption excess for our
entire sample, the middle and right panels of Fig.~\ref{fig:two} show
- for objects where such an absorption excess is detected - how the
position and width of our best-fitting NaD profile compares to that of
the stellar line-of-sight velocity distribution.
In these $V_{\rm NaD}-V_{\rm stars}$ vs. $\sigma_{\rm NaD}-\sigma_{\rm
  stars}$ diagrams we can identify objects with a) possible NaD
outflows where $\Delta V \la-100 \,\rm km\,s^{-1}$, b) relaxed dusty
disks where $\Delta V \sim 0$ but $\Delta \sigma < 0$, corresponding
to narrower NaD interstellar absorption lines compared to stellar
photospheric features, c) possible mild inflows due to bars or
unsettled dust lanes where $\Delta \sigma < 0$ and $\Delta V > 0$, and
finally
d) systems that systematically show $\Delta \sigma > 0$, $\Delta
V\sim0$ and no evidence for an interstellar medium as their {\tt
  GandALF\/} fits imply little or no need for reddening by dust
(Fig.~\ref{fig:two}, right panel) and only weak or no nebular
emission.
The NaD excess of these last objects is almost certainly due to
template-mismatch, where the large width of such an excess is likely
driven by an enhanced presence of cool 
%
%
stars and the strong pressure-broadened wings of their NaD
photospheric lines.

%
%
There are 23 objects with possible NaD outflows in our sample, most of
which (74\%) exhibit either SF or TO central activity,
consistent with SDSS studies on starburst-driven outflows
\citep[e.g.,][]{Che10} and possibly with an evolution from starburst
to quenched galaxies with star formation decreasing along this path as
AGN emerge \citep[e.g.,][]{Yes14}, respectively. On the other hand,
there is little evidence for NaD outflows in galaxies harbouring
Seyfert nuclei or LINER emission.
The few objects with little or no nebular emission that show evidence
of NaD outflow in Fig.\ref{fig:two} (including a spectacular case with
$\Delta V \sim -1000\,\rm km\,s^{-1}$ shown by the left-pointing
arrow) are most likely SF or TO where the \Oiii\ line remained
undetected due to large amounts of reddening.
%
%

\subsection{mJIVE-20 Detected Objects}
\label{sec:results_mJIVE}

Around 23\% of our FIRST-selected sample of galaxies (103 objects)
show unequivocal signatures of AGN activity through to the detection
of a compact radio core in the mJIVE-20 images.
These are generally (79\% of the cases) objects that show only weak or
no nebular emission, or which are otherwise evenly distributed between
the regions of the \Oiii/\Hb\ vs. \Nii/\Ha\ BPT diagram that include
transition objects, Seyfert nuclei and galaxies with LINER emission,
consistent with the presence of an AGN as signalled by the finding of
a compact radio core.
In fact, central SF activity in our VLBI-detected objects is likely to
be rather limited given that most of their FIRST flux is unresolved
(90 objects have more than 80\% of their total FIRST flux density in
an unresolved component) and that generally (86 objects, or 83\%) more
than half of this compact emission is found in the VLBI component.
Furthermore, VLBI-detected galaxies reside primarily on the
red-sequence of passively-evolving system in the UV-optical
color-magnitude diagram \citep{Kav15a}, which is sensitive to even
small fractions of young stars \citep{Yi05,Kav07}.

Among the VLBI-detected objects, 20 ($\sim19\%$) show an NaD
absorption excess, and these are predominantly ETGs with little or
weak nebular emission (Fig.~\ref{fig:three}). For a third of these
galaxies, however, such an excess is most likely induced by
template-mismatch given their position in the $V_{\rm NaD}-V_{\rm
  stars}$ vs. $\sigma_{\rm NaD}-\sigma_{\rm stars}$ diagram and their
lack of reddening by dust.
The remaining objects, a dozen at best, show no evidence of outflows
and are instead consistent with the presence of relaxed dusty disks or
slight inflows, possibly indicating unsettled dust-lanes.
Thus, only a minority of nearby radio AGN is found in galaxies
where NaD absorption indicates the presence of cold-gas, and in no
systems do we find evidence for such a medium to be outflowing.

This result contrasts with the findings of \citet{Leh11}, who detected
an NaD excess in approximately 1/3 of their sample of radio-loud ETGs
and with NaD profiles showing little or no kinematic offset ($\Delta
V\sim-50\,\rm km\,s^{-1}$) but very broad profiles ($\sigma_{\rm
  NaD}\sim500\,\rm km\,s^{-1}$).
Part of this discrepancy may be due to differences in sample
selection.
\citeauthor{Leh11} analysed a sample of 700 radio-loud ETGs at
$z<0.2$, with FIRST 1.4 GHz flux densities above 40 mJy and resolved
jet morphologies. This selection would naturally sample high-density
environments such as galaxy clusters where a hot-gas medium is
available for the AGN to work against and produce detectable radio
lobes.
On the other hand, the mJIVE-20 survey is capable of detecting compact
radio AGN regardless of local environment and thus probes AGNs in the
general galaxy population that largely inhabits relatively
low-density environments \citep{Kav15a}.
The working of radio AGN activity may thus be different in these two
samples, being directly fuelled by the cooling of hot gas in cluster
centres in the former \citep{Har07,Sha08} while being triggered by
minor mergers in the latter \citep{Bes12,Kav15a}.
Additionally, it may be that also some of the NaD excess detected
by \citeauthor{Leh11} is induced by template-mismatch, in particular
given than the limitations of stellar-population models are most
evident in massive ETGs such as central cluster galaxies.
In fact, it is only in the most massive ETGs of our sample (with
$\sigma_{\rm star}\sim250\,\rm km\,s^{-1}$ ) that we find NaD excess
due to template-mismatch, with NaD profiles as wide (up to
$\sigma_{\rm NaD}\sim500\,\rm km\,s^{-1}$) as those observed by
\citeauthor{Leh11}.

\section{Conclusions}
\label{sec:conclusions}

Using the NaD interstellar absorption as a tracer of neutral gas we
have looked for cold-gas outflows in a sample of 456 nearby galaxies
with known central radio-continuum emission from the FIRST survey and
for which we could unambiguously ascertain the presence of nuclear
radio activity using our mJIVE-20 radio imaging at milli-arcsecond
scales.
%
%
VLBI-detected radio cores are found in approximately 23\% of our
sample (103 objects) and occur predominantly in early-type galaxies
with little or no nebular emission in their SDSS spectra and which are
dominated by old stellar populations.
Only in a dozen objects such radio AGN activity is accompanied by
interstellar NaD absorption, but the neutral gas traced by the NaD
lines never appears to be outflowing.
%
%
On the other hand, interstellar NaD absorption is detected in 1/3 of
the objects that do not possess a VLBI radio core (in 122 out 353
objects), consistent with the presence of a conspicuous gaseous medium
where most of the time star formation contributes significantly to
their central FIRST radio emission. Only in such objects - without a
radio AGN - do we find evidence of cold-gas outflows, and these
generally arise in spiral galaxies with central SF or TO nebular
activity.

These findings reinforce the picture in which radio AGN activity
occurs most often in systems where the gas reservoir has already been
significantly depleted.
Interestingly, across our entire sample cold-gas outflows occur only
in objects dominated by central starburst or composite
AGN/star-formation activity whereas optical Seyfert AGN activity does
not seem capable of driving such winds.
This result suggests either that only supernovae feedback can drive
galactic winds in the objects that we have studied or that the role of
optical AGN activity in this respect is limited to a phase accompanied
by significant star-formation activity.

\section*{Acknowledgements} 
SSS thanks the Australian Research Council for an Early Career
Fellowship, DE130101399. MS and JT acknowledge the support of a
SEPnet summer placement, which enabled the initial steps of
this work. MS dedicates this paper to his daughter Morgana Rose.

\label{lastpage}

\begin{thebibliography}{}

\bibitem[\protect\citeauthoryear{Abazajian et al.}{2009}]{Aba09}
  Abazajian K.~N., et al., 2009, ApJS, 182, 543

\bibitem[\protect\citeauthoryear{Baldwin, Phillips \&
    Terlevich}{1981}]{Bal81} Baldwin, J. A., Phillips, M. M.,
  Terlevich, R. 1981, PASP, 93, 5

\bibitem[\protect\citeauthoryear{Becker, White \&
    Helfand}{1995}]{Bec95} Becker R.~H., White R.~L., Helfand D.~J.,
  1995, ApJ, 450, 559

\bibitem[\protect\citeauthoryear{Best \& Heckman}{2012}]{Bes12} Best
  P.~N., Heckman T.~M., 2012, MNRAS, 421, 1569

\bibitem[\protect\citeauthoryear{Cappellari \& Emsellem}{2004}]{Cap04}
  Cappellari M., Emsellem E., 2004, PASP, 116, 138

\bibitem[\protect\citeauthoryear{Chen et al.}{2010}]{Che10} Chen
  Y.-M., Tremonti C.~A., Heckman T.~M., Kauffmann G., Weiner B.~J.,
  Brinchmann J., Wang J., 2010, AJ, 140, 445

\bibitem[\protect\citeauthoryear{Croton et al.}{2006}]{Cro06} Croton
  D.~J., et al., 2006, MNRAS, 365, 11

\bibitem[\protect\citeauthoryear{Deller \& Middelberg}{2014}]{Del14}
  Deller A.~T., Middelberg E., 2014, AJ, 147, 14

\bibitem[\protect\citeauthoryear{Fabian}{2012}]{Fab12} Fabian A.~C.,
  2012, ARA\&A, 50, 455

\bibitem[\protect\citeauthoryear{Jeong et al.}{2007}]{Jeo07} Jeong H.,
  Bureau M., Yi S.~K., Krajnovi{\'c} D., Davies R.~L., 2007, MNRAS,
  376, 1021

\bibitem[\protect\citeauthoryear{Jeong et al.}{2013}]{Jeo13} Jeong H.,
  Yi S.~K., Kyeong J., Sarzi M., Sung E.-C., Oh K., 2013, ApJS, 208, 7

\bibitem[\protect\citeauthoryear{Hardcastle, Evans, \&
    Croston}{2007}]{Har07} Hardcastle M.~J., Evans D.~A., Croston
  J.~H., 2007, MNRAS, 376, 1849

\bibitem[\protect\citeauthoryear{Kauffmann et al.}{2003}]{Kau03}
  Kauffmann G., et al., 2003, MNRAS, 346, 1055

\bibitem[\protect\citeauthoryear{Kaviraj et al.}{2007}]{Kav07} Kaviraj
  S., et al., 2007, ApJS, 173, 619

\bibitem[\protect\citeauthoryear{Kaviraj}{2009}]{Kav09} Kaviraj S.,
  2009, MNRAS, 394, 1167

\bibitem[\protect\citeauthoryear{Kaviraj et al.}{2015a}]{Kav15a} Kaviraj
  S., Shabala S.~S., Deller A.~T., Middelberg E., 2015a, MNRAS, in
  press, arXiv:1412.5602

\bibitem[\protect\citeauthoryear{Kaviraj et al.}{2015b}]{Kav15b} Kaviraj
  S., Shabala S.~S., Deller A.~T., Middelberg E., 2015b, MNRAS, 452,
  774

\bibitem[\protect\citeauthoryear{Kewley et al.}{2001}]{Kew01} Kewley
  L.~J., Dopita M.~A., Sutherland R.~S., Heisler C.~A., Trevena J.,
  2001, ApJ, 556, 121

\bibitem[\protect\citeauthoryear{Lehnert et al.}{2011}]{Leh11} Lehnert
  M.~D., Tasse C., Nesvadba N.~P.~H., Best P.~N., van Driel W., 2011,
  A\&A, 532, L3

\bibitem[\protect\citeauthoryear{Nesvadba et al.}{2008}]{Nev08}
  Nesvadba N.~P.~H., Lehnert M.~D., De Breuck C., Gilbert A.~M., van
  Breugel W., 2008, A\&A, 491, 407

\bibitem[\protect\citeauthoryear{Nyland et al.}{2013}]{Nyl13} Nyland
  K., et al., 2013, ApJ, 779, 173

\bibitem[\protect\citeauthoryear{Oh et al.}{2011}]{Oh11} Oh K., Sarzi
  M., Schawinski K., Yi S.~K., 2011, ApJS, 195, 13

\bibitem[\protect\citeauthoryear{Park, Jeong and Yi}{2015}]{Par15}
  Park, J., Jeong H., Yi S.~K., 2015, ApJ, in press, arXiv:1507.03342

\bibitem[\protect\citeauthoryear{Rupke, Veilleux, \&
    Sanders}{2005}]{Rup05} Rupke D.~S., Veilleux S., Sanders D.~B.,
  2005, ApJS, 160, 87

\bibitem[\protect\citeauthoryear{Sarzi et al.}{2006}]{Sar06} Sarzi M.,
  et al., 2006, MNRAS, 366, 1151

\bibitem[\protect\citeauthoryear{Sato et al.}{2009}]{Sat09} Sato T.,
  Martin C.~L., Noeske K.~G., Koo D.~C., Lotz J.~M., 2009, ApJ, 696,
  214

\bibitem[\protect\citeauthoryear{Schawinski et al.}{2007}]{Sch07}
  Schawinski K., Thomas D., Sarzi M., Maraston C., Kaviraj S., Joo
  S.-J., Yi S.~K., Silk J., 2007, MNRAS, 382, 1415

\bibitem[\protect\citeauthoryear{Shabala et al.}{2008}]{Sha08} Shabala
  S.~S., Ash S., Alexander P., Riley J.~M., 2008, MNRAS, 388, 625

\bibitem[\protect\citeauthoryear{Shabala \& Alexander}{2009}]{Sha09}
  Shabala S., Alexander P., 2009, ApJ, 699, 525

\bibitem[\protect\citeauthoryear{Shabala et al.}{2012}]{Sha12} Shabala
  S.~S., et al., 2012, MNRAS, 423, 59

\bibitem[\protect\citeauthoryear{Silk \& Rees}{1998}]{Sil98} Silk J.,
  Rees M.~J., 1998, A\&A, 331, L1

\bibitem[\protect\citeauthoryear{Veilleux \& Osterbrock}{1987}]{Vei87}
  Veilleux, S., Osterbrock, D. E. 1987, ApJS, 63, 295

\bibitem[\protect\citeauthoryear{Wild, Heckman \&
    Charlot}{2010}]{Wil10} Wild V., Heckman T., Charlot S., 2010,
  MNRAS, 405, 933

\bibitem[\protect\citeauthoryear{Yesuf et al.}{2014}]{Yes14} Yesuf
  H.~M., Faber S.~M., Trump J.~R., Koo D.~C., Fang J.~J., Liu F.~S.,
  Wild V., Hayward C.`C., 2014, ApJ, 792, 84

\bibitem[\protect\citeauthoryear{Yi et al.}{2005}]{Yi05} Yi S.~K. et
  al. 2005, ApJ, 619, L111

\end{thebibliography}
\end{document}